\documentclass[cits,a4paper]{PoS}
\usepackage{amsmath,amssymb}
\usepackage[tight,TABTOPCAP]{subfigure}
\usepackage{graphicx}
\usepackage{grffile}

\DeclareRobustCommand{\amuhlo}{$a_\mu^{\mathrm{HLO}}$}

\title{Study of the anomalous magnetic moment of the muon computed from the Adler function}

\ShortTitle{The anomalous magnetic moment of the muon from the Adler function}
     
\author{\vspace{-0.75cm}\phantom{a} \hfill \textnormal{MITP/14-090}\newline\newline
Michele Della Morte$^1$, Anthony Francis$^2$, Gregorio
Herdoiza$^4$, \speaker{Hanno Horch}$^3$, Benjamin J\"ager$^{5}$, Andreas J\"uttner$^6$, Harvey
Meyer$^3$, Hartmut Wittig$^{2,3}$\\
$^{1}$ CP3-Origins \& Danish IAS, University of Southern Denmark\\
		 Campusvej 55, DK-5230 Odense M, Denmark and IFIC (CSIC)\\
		 Calle Catedr\'{a}tico Jos\'{e} Beltr\'{a}n, 2. E-46980, Paterna, Spain\\
$^{2}$Helmholtz Institute Mainz, Johannes Gutenberg Universit\"at Mainz,
       55099 Mainz, Germany \\
 $^{3}$PRISMA Cluster of Excellence, Institut f\"ur Kernphysik, Johannes
	Gutenberg Universit\"at Mainz, 55099 Mainz, Germany\\
$^{4}$  Instituto de F\'isica Te\'orica UAM/CSIC and
Departamento de F\'isica Te\'orica,\\
Universidad Aut\'onoma de Madrid, Cantoblanco, E-28049
Madrid, Spain\\
$^{5}$ Department of Physics, College of Science, Swansea University, SA2 8PP, Swansea, UK\\
$^{6}$ School of Physics and Astronomy\\
	   University of Southampton, UK\\  
E-mail:\email{horch@kph.uni-mainz.de}, \email{dellamor@ific.uv.es}
	   \email{francis@kph.uni-mainz.de},
       \email{gregorio.herdoiza@uam.es},
       \email{B.Jaeger@swansea.ac.uk},
       \email{a.juttner@soton.ac.uk},
       \email{meyerh@kph.uni-mainz.de}, \email{wittig@kph.uni-mainz.de}\\}

\abstract{We compute the Adler function on the lattice from vacuum polarization data with twisted
boundary conditions using numerical derivatives. The study is based on CLS ensembles with two flavours of
$O(a)$ improved Wilson fermions. We extrapolate the lattice data for the Adler function to the
continuum limit and to the physical pion mass and analyze its dependence on the momentum transfer.
We discuss the application of this method to the extraction of the $u,d$ contribution to \amuhlo.}

\FullConference{The 32nd International Symposium on Lattice Field Theory\\
                 23-28 June, 2014\\
                 Columbia University New York, NY}

\begin{document}

\section{Introduction}
The anomalous magnetic moment of the muon is given by $a_\mu\equiv(g_\mu-2)/2$, where
$g_\mu$ is the gyromagnetic factor of the muon. There has been a persistent deviation for the
results obtained from theory and experiment \cite{PDG2014} which is currently $\sim 3.6\sigma$,
\begin{align*}
a_\mu^{exp}&=116\,592\,091(54)(33) \cdot 10^{-11},\\		
a_\mu^{th} &=116\,591\,803(01)(42)(26)\cdot 10^{-11}.	
\end{align*}
This discrepancy may hint at new physical processes beyond the Standard Model. A new experiment at
Fermilab aims at reducing the experimental error by a factor of four \cite{Carey:2009zzb}. To fully
exploit this level of accuracy, it is mandatory to also increase the theoretical precision. The error budget of the
theoretical result is dominated by QCD effects. The error due to the leading order ($\alpha^2$)
contribution is  $42\cdot 10^{-11}$, and for the next to leading order ($\alpha^3$) contribution it
is $26\cdot 10^{-11}$, while effects from weak interactions and QED are of the order of $10^{-11}$
and $10^{-13}$, respectively. Lattice QCD allows us to determine the leading hadronic contribution
to the anomalous magnetic moment of the muon $\left(a_\mu^{\mathrm{HLO}}\right)$ from the hadronic
vacuum polarization (VP). The VP tensor is given by
\begin{align}
\Pi_{\mu\nu}(Q^2)&=\int d^4x e^{iQx}\left<J_\mu(x)J_\nu(0)\right>, \quad
\Pi_{\mu\nu}(Q^2)=\left(Q_\mu Q_\nu - \delta_{\mu\nu}Q^2\right)\Pi(Q^2),
\label{vpfdetermation}
\end{align}
where the second relation follows from Euclidean invariance and current conservation. In order to
perform the convolution integral \cite{deRafael:1993za, Blum:2002ii}, which yields \amuhlo, one
needs the renormalized VP, $\hat{\Pi}(Q^2)=4\pi^2(\Pi(Q^2)-\Pi(0))$, i.e.
\begin{align}
 a_\mu^\mathrm{HLO}&=\left(\frac{\alpha}{\pi}\right)^2\int_{0}^{\infty}
 dQ^2f(Q^2)\hat{\Pi}(Q^2), 
 \label{amustandardmethod}\\
 f(Q^2)&=\frac{m_\mu^2Q^2Z^3(1-Q^2Z)}{1+m_\mu^2 Q^2Z^2}, \quad
 Z=-\frac{Q^2-\sqrt{Q^4+4m_\mu^2Q^2}}{2m_\mu^2 Q^2},\nonumber 
 \end{align}
to obtain \amuhlo. This has been studied by a number of lattice collaborations \cite{aubinblum07,
Boyle:2011hu, DellaMorte:2011aa, Burger:2013jya}. This work is a continuation of
\cite{Horch:2013lla}, and is part of the $g-2$ project in Mainz, cf. \cite{DellaMorte:2011aa,
 Herdoiza:2014jta, Francis:2013fzp, francis_g-2_lat_2014}.
The largest contribution to the integral in eq. \eqref{amustandardmethod} is dominated by the low
$Q^2$ region around $m_\mu^2$, which is also a difficult region to probe on the lattice, since
statistical fluctuations rapidly increase as $Q^2\rightarrow 0$. While we cannot circumvent the need
for precise data in this region, we can remove the need to extrapolate to $\Pi(Q^2 = 0)$. This can
be achieved by studying the Adler function \cite{Adler:1974gd}, defined as
\begin{align}
D(Q^2)&=12 \pi^2Q^2 \frac{d\Pi(Q^2)}{d(Q^2)}.
\label{adlerfunction}
\end{align}
Making use of partially twisted boundary conditions
\cite{Sachrajda:2004mi,Bedaque:2004ax,deDivitiis:2004kq}, we apply two procedures to determine the
numerical derivative of $\Pi(Q^2)$ and to check for inherent systematic effects.

\section{Lattice setup and determination of the Adler function}
Our study is based on CLS \cite{CLSwiki,CLSalgo,CLSpos} ensembles with $N_f=2$ $O(a)$-improved
Wilson fermions, cf. table \ref{clsensembles}, where partially twisted boundary
conditions are imposed on the valence quarks.

\begin{table}[ht!]
\centering
\begin{tabular}{|c|c|c|c|c|c|c|c|c|c|}
\hline
 Label  & V$/a^4$ &$\beta$ &   $a$[fm] & $m_\pi$[MeV] & $m_\pi L$ & $N_{\rm cnfg}$ & $N_{\rm meas}$
 \\
\hline
A3 & $64\times 32^3$&   5.20 & 0.079 & 473 & 6.0 & 251 & 1004\\
A4 & $64\times 32^3$&   5.20 & 0.079 & 363 & 4.7 & 400 & 1600\\
A5 & $64\times 32^3$&   5.20 & 0.079 & 312 & 4.0 & 251 & 1004\\
B6 & $96\times 48^3$&   5.20 & 0.079 & 267 & 5.1 & 306 & 1224\\
\hline
E5 & $64\times 32^3$&   5.30 & 0.063 & 456 & 4.7 & 1000 & 4000\\
F6 & $96\times 48^3$&   5.30 & 0.063 & 325 & 5.0 &  300 & 1200\\
F7 & $96\times 48^3$&   5.30 & 0.063 & 277 & 4.2 &  250 & 1000\\
G8 & $128\times64^3$&   5.30 & 0.063 & 193 & 4.0 &  205 &  820\\
\hline
N5 & $96\times 48^3$&   5.50 & 0.050 & 430 & 5.2 & 347 & 1392\\
N6 & $96\times 48^3$&   5.50 & 0.050 & 340 & 4.1 & 559 & 2236\\
O7 & $128\times64^3$&   5.50 & 0.050 & 261 & 4.4 & 138 &  552\\
\hline
\end{tabular}
\caption{The CLS ensembles used in this study with the lattice spacing from \cite{latticespacing}
and the number of configurations $N_{\rm cnfg}$ and measurements $N_{\rm meas}$.}
\label{clsensembles}
\end{table}

From eqs. \eqref{vpfdetermation} one can determine the VP function on the lattice. As an example we
show the measurement for the ensemble F7 with $a=0.063\,$fm and $m_\pi=277\,$MeV in fig.$\,$1.
To compute the Adler function we apply a linear interpolation within a certain interval of the
data. The interpolation interval must be chosen with care: On the one hand, local fluctuations in
the VP data may spoil a reliable determination of the numerical derivative if the interval is chosen
too small. At the same time choosing too large a fit window cannot model the curvature accurately.
We developed two procedures which use different criteria to select the best fit window at each $Q^2$
value.\newline
The first procedure (proc.~I) uses linear fits $\Pi_{fit}(Q^2)=a_I+b_IQ^2$ with
several fit windows $\epsilon\,\in\,[0.1,1.0]$GeV$^2$ at each momentum transfer. We
select the best fit $\Pi_{fit}(Q^2)$ from the region where $b_I$ is stable with respect to the fit
window size $\epsilon$. The Adler function is then given by $b_IQ^2$. The second procedure
(proc.~II) also uses linear fits $\Pi^{(l)}_{fit}(Q^2)=a_{II}+b_{II}\log(Q^2)$ and quadratic fits
$\Pi^{(q)}_{fit}(Q^2)=a^\prime_{II}+b^\prime_{II}\log(Q^2)+c^\prime_{II}\log(Q^2)^2$ with same fit
windows $\epsilon\,\in\,[0.1,1.0]$GeV$^2$ as for proc.~I at each momentum transfer. We apply
cuts to the curvature $c^\prime_{II}$, and to the $\chi^2/$dof. By requiring $b_{II}\simeq
b^\prime_{II}$ we choose our final results from the remaining set of fits. The factor $b_{II}$ then
determines the Adler function.\newline 
As a cross check we fit a Pad\'{e}-[1,2] ansatz to the VP data and compute its derivative. The
results for these procedures are shown on the right in fig.$\,$1. We find that the three discussed
procedures match within errors across the considered region of $Q^2$.

\begin{figure}[ht!]
  \centering 
  \includegraphics{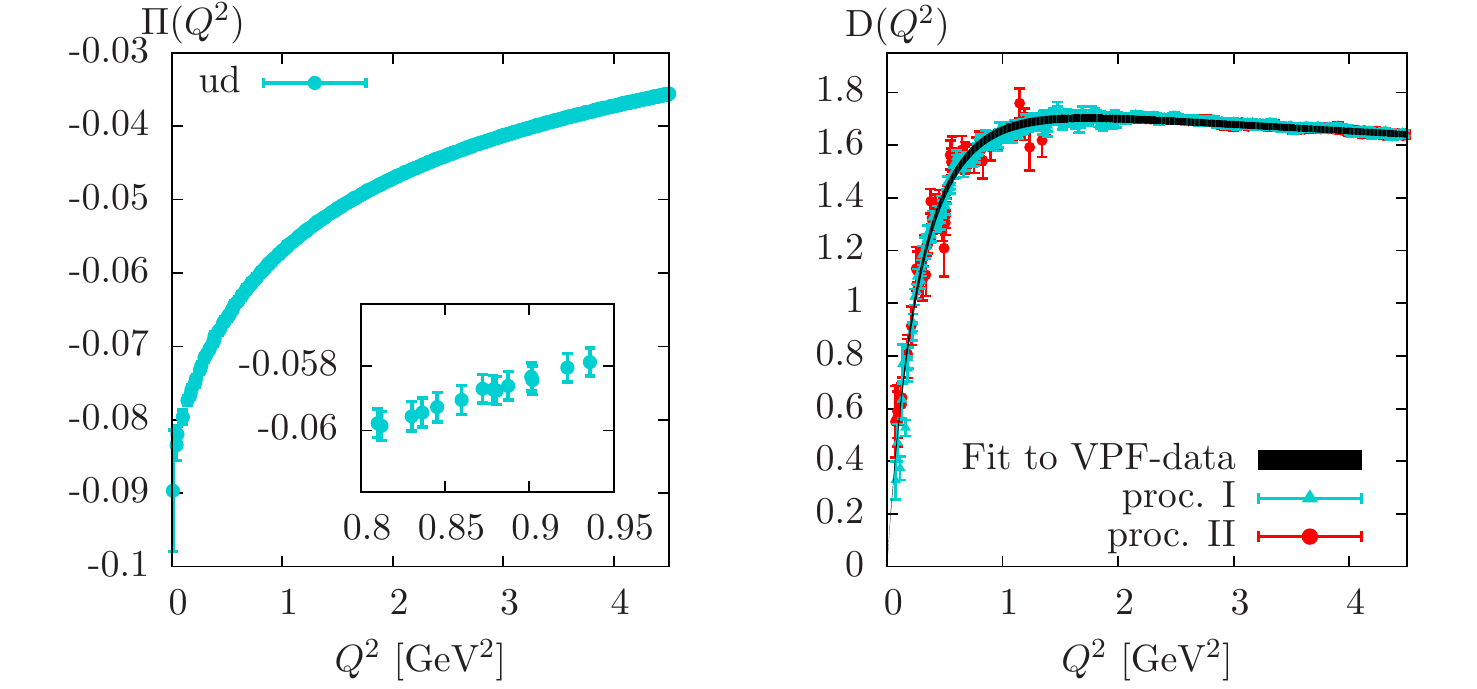}
\label{f7vpfadler}
\caption{Left: The VP function on F7 with $a=0.063\,$fm and
$m_\pi=277\,$MeV. Right: Results for the Adler function using the two numerical procedures
discussed in the text. The derivative of fitting the VP to a Pad\'{e}[1,2]-Ansatz is shown as a
cross check.}
\end{figure}

\section{The Adler function in the continuum at the physical point}
In order to take the continuum limit and extrapolate the Adler function to the physical pion mass we
perform combined fits of the type
\begin{align}
D_{ijk}(Q^2)&=A_{[ij]}(Q^2)(1 + B_k(a,Q) + C(m_\pi,Q^2)),
\end{align}
where the indices $i,j,k$ are labels for the different fit ans\"atze that we considered. The factor
$A$ models the $Q^2$ dependence, $B$ describes lattice artifacts, and $C$ the light quark mass
dependence. We use two ans\"atze for the momentum dependence
\begin{align}
A_{[12]}(Q^2)&=Q^2\left(\frac{a_1b_1}{\left(b_1+Q^2\right)^2} +
	\frac{a_2 b_2}{\left(b_2+Q^2\right)^2}\right),\quad
A_{[22]}(Q^2)=Q^2\left(\frac{a_1 b_1}{\left(b_1+Q^2\right)^2} +
	\frac{a_2 b_2}{\left(b_2+Q^2\right)^2}+a_0\right),
\end{align}
which are based on derivatives of the Pad\'{e} ans\"atze $[1,2]$ and $[2,2]$, cf.
Refs. \cite{DellaMorte:2011aa, Aubin:2012me}. To describe the lattice spacing dependence we consider
two possibilities
\begin{align}
		B_1(a,Q)&=c_1(aQ) + c_2 (4\pi f_{\rm K} a),\quad
		B_2(a,Q)=c_1\left(aQ\right)^2 + c_2 \left(4\pi f_{\rm K} a\right)^2,
\end{align}
where the term proportional to $c_1$ describes lattice artifacts which depend on the momentum
transfer. The factor $4\pi f_K$ in the momentum-independent term merely renders $c_2$ dimensionless.
Furthermore, we explore two alternatives, namely that the leading lattice artifacts are either
$\rm O(a)$ or $\rm O(a^2)$. To describe the light quark mass dependence we use
\begin{align}
	C(m_\pi,Q^2)&=d_1\frac{m_\pi^2-\left(m_\pi^{\rm phys}\right)^2}{d_2+Q^2}.
\end{align}  
In fig.$\,$2 we show the result of an uncorrelated fit for the ansatz $D_{222}$
including all available ensembles, cf. table \ref{clsensembles}, for the Adler function obtained
using proc.~II. The top left shows the Adler function data and the fit result as dotted, dashed,
and solid lines for ensembles with different lattice spacings with a roughly constant pion mass of
$m_\pi\approx 270\,$MeV. The continuous filled black curve is the extrapolation to the continuum at
this pion mass. We find that lattice artifacts are small for the low $Q^2$ region, and grow for
larger values of $Q^2$. This is further underlined by the plot on the top right where the slopes at
$Q^2=1.0$ and $3.0\,$GeV$^2$ are compared.\newline
 The bottom left of fig.\,2 shows the result for the Adler function data for a fixed lattice
spacing and the evaluation of the fit function for each pion mass shown as dashed, dotted, and solid
lines. The continuous filled black curve represents the extrapolation to the physical pion mass. We
observe that the pion mass dependence is mild at large $Q^2$ and becomes more and more significant
when decreasing $Q^2$. The bottom right shows the dependence of the pion mass for two fixed momentum
transfers. The vertical black line represents the physical pion mass.

 \begin{figure}[t!]
  \centering 
  \includegraphics{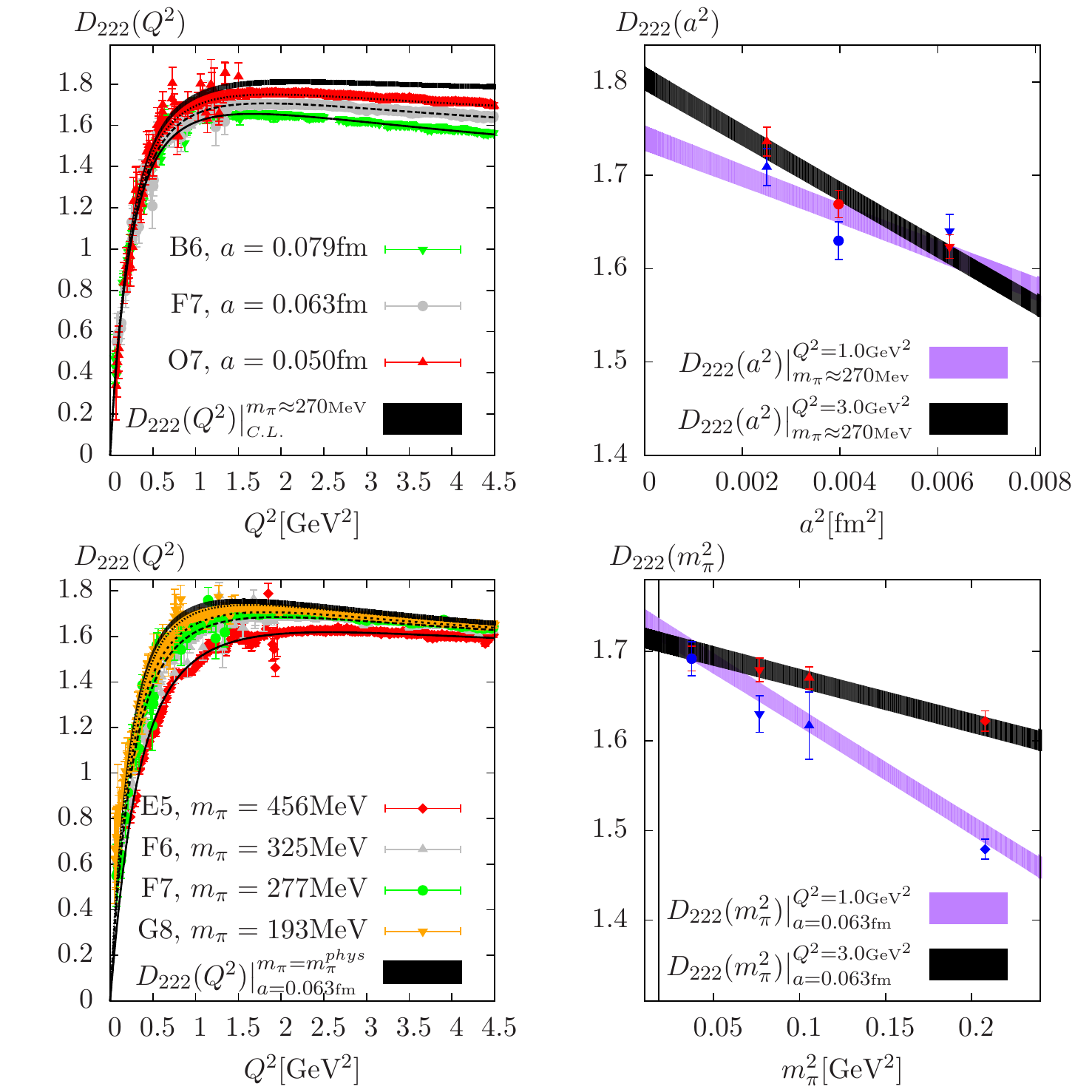}
\label{combinedfitplot}
\caption{Illustration of an uncorrelated fit to the Adler function including the complete set of
ensembles in table \protect\ref{clsensembles} for the Adler function obtained using proc.~II with
the ansatz $D_{222}$, i.e. a derivative of a Pad\'{e} [2,2] ansatz with $O(a^2)$ effects. C.L.
refers to the continuum limit. The results are discussed in the text. Top left: Comparison of the
Adler function for the available lattice spacings and constant pion mass including the continuum
extrapolation. Top right: Lattice spacing dependence at fixed $Q^2=1.0$ and 3.0\,GeV$^2$. Bottom
left: Comparison of the pion mass dependence fore a fixed lattice spacing including the
extrapolation to the physical pion mass. Bottom right: Pion mass dependence at fixed $Q^2=1.0$ and
3.0\,GeV$^2$.}
\end{figure}

\section{Determination of \amuhlo}
\amuhlo can be obtained via the integral \cite{Lautrup:1971jf, Knecht:2003kc}
\begin{align}
a_\mu^\mathrm{HLO}&=\int_{0}^{1}dxg(x)D\left(\frac{x^2m_\mu^2}{1-x}\right),\quad
g(x)=\frac{\alpha^2}{6\pi^2}\frac{(1-x)(2-x)}{x}
\label{amuadler}
\end{align}
where we used the substitution $Q^2\rightarrow\frac{x^2m_\mu^2}{1-x}$ and inserted the Adler
function $D(Q^2)$ in the continuum limit at the physical pion mass, obtained from the
combined fit. The left of fig.$\,$3 illustrates the integrand of eq. \eqref{amuadler} for the case
of the ensemble F7, $m_\pi=277\,{\rm MeV}$ and $a=0.063\,{\rm fm}.$ The expected large contribution
to $a_\mu^{\rm HLO}$ from momenta in the neighborhood of the muon mass (denoted by the vertical line
in fig.$\,$3) can be probed by the use of Pad\'{e} approximants of different orders. The right of
fig.$\,$3 shows the results obtained for \amuhlo using the different fit functions discussed in the
previous section to the Adler function data obtained via proc.~II. Lattice artifacts are well
controlled, since similar results are obtained for \amuhlo when using $O(a)$ and $O(a^2)$ ans\"atze.
When all ensembles are included in the combined fit we observe a deviation for the two orders of the
Pad\'{e} ans\"atze that we considered. Removing the ensembles with $m_\pi > 400$\,MeV we find the
deviation is reduced and the results agree within errors, which is in part due to an increase of the
latter.

\begin{figure}[ht!]
  \centering 
  \includegraphics{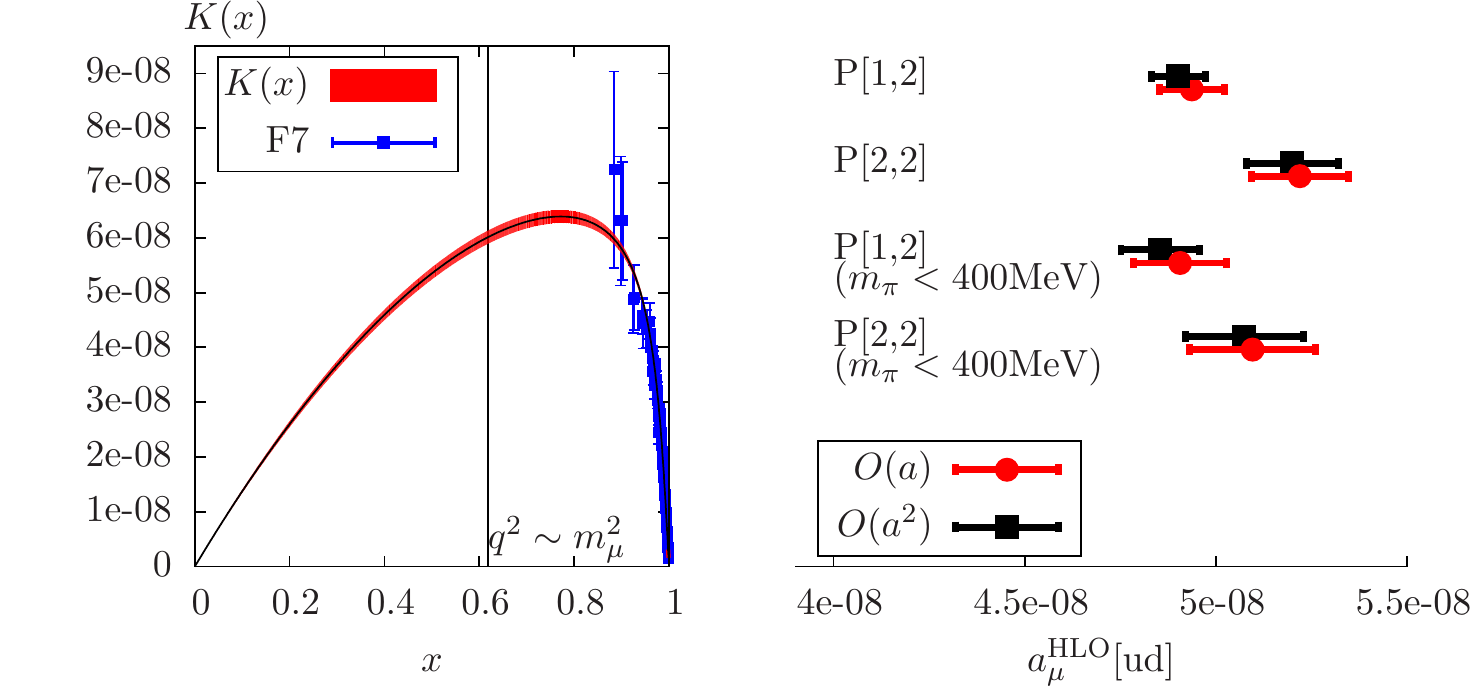}
\label{amulight}
\caption{Left: Plot of the integrand of eq. \protect\eqref{amuadler} with
$K(x)=g(x)D_{222}\left((x^2m_\mu^2)/(1-x)\right)$ evaluated at the parameters of the F7
ensemble, in comparison with the data of the Adler function obtained via proc.~II multiplied by
$g(x)$. Right: Preliminary results of the $u,d$ contribution to $a_\mu^{\rm HLO}$ from uncorrelated
fits to the Adler function data obtained via proc.~II.}
\end{figure}

\section{Conclusions}
We presented two numerical procedures to determine the Adler function from $\Pi(Q^2)$ through
lattice calculations. Both procedures match within errors for a large range in $Q^2$. After
extrapolating the Adler function to the continuum and to the physical pion mass we use a kernel
representation based on the Adler function to compute \amuhlo. We showed that lattice artifacts for
\amuhlo  are small. We observe the results are sensitve to the chiral approach of the data to the
physical point and the $Q^2$-dependence of the Pad\'{e} approximants. We plan to investigate
different ans\"atze to further study these effects. Furthermore, we will also extend
this study to the strange and charm quark contribution to \amuhlo.\newline 
 \textsc{Acknowledgements}: We thank Andreas Nyffeler for pointing out the kernel
 function in eq. \eqref{amuadler} to us. Our calculations were performed on the ``Wilson'' and
 ``Clover'' HPC Clusters at the Institute for Nuclear Physics, University of Mainz. We thank Dalibor
 Djukanovic and Christian Seiwerth for technical support. This research has been supported in part
 by the DFG via the SFB~1044. G.H. acknowledges support by the the Spanish MINECO through the
 Ram\'on y Cajal Programme and through the project FPA2012-31686 and by the Centro de excelencia
 Severo Ochoa Program SEV-2012-0249. This work was granted access to the HPC resources of the Gauss
 Center for Supercomputing at Forschungzentrum J\"ulich, Germany, made available within the
 Distributed European Computing Initiative by the PRACE-2IP, receiving funding from the European
 Community's Seventh Framework Programme (FP7/2007-2013) under grant agreement RI-283493 (project
 PRA039) and ERC grant agreement No 279757.


\begin{thebibliography}{99}
\setlength{\itemsep}{-0.2mm}
\bibitem{PDG2014}
  K.~A.~Olive {\it et al.}  [Particle Data Group Collaboration],
  Chin.\ Phys.\ C {\bf 38} (2014) 090001.
  
\bibitem{Carey:2009zzb}
  R.~M.~Carey, K.~R.~Lynch, J.~P.~Miller, B.~L.~Roberts, W.~M.~Morse, Y.~K.~Semertzides, V.~P.~Druzhinin and B.~I.~Khazin {\it et al.},
  FERMILAB-PROPOSAL-0989.


\bibitem{deRafael:1993za}
  E.~de Rafael,
  Phys.\ Lett.\ B {\bf 322} (1994) 239, hep-ph/9311316.


\bibitem{Blum:2002ii}
  T.~Blum,
  Phys.\ Rev.\ Lett.\  {\bf 91} (2003) 052001, hep-lat/0212018.

\bibitem{aubinblum07}
  C.~Aubin and T.~Blum,
  Phys.\ Rev.\ D {\bf 75} (2007) 114502, hep-lat/0608011.
  
\bibitem{Boyle:2011hu}
  P.~Boyle, L.~Del Debbio, E.~Kerrane and J.~Zanotti,
  Phys.\ Rev.\ D {\bf 85} (2012) 074504, arXiv:1107.1497.
\bibitem{DellaMorte:2011aa}
  M.~Della Morte, B.~J\"ager, A.~J\"uttner and H.~Wittig,
  JHEP {\bf 1203} (2012) 055, arXiv:1112.2894.
\bibitem{Burger:2013jya}
  F.~Burger {\it et al.}  [ETM Collaboration],
  JHEP {\bf 1402} (2014) 099, arXiv:1308.4327.

\bibitem{Horch:2013lla}
  H.~Horch {\it et al.} 
  PoS LATTICE {\bf 2013} (2013) 304, arXiv:1311.6975.
  
\bibitem{Herdoiza:2014jta}
  G.~Herdo\'iza, H.~Horch, B.~J\"ager and H.~Wittig,
  PoS LATTICE {\bf 2013} (2014) 444.
  
\bibitem{Francis:2013fzp}
  A.~Francis, B.~J\"ager, H.~B.~Meyer and H.~Wittig,
  Phys.\ Rev.\ D {\bf 88} (2013) 054502, arXiv:1306.2532.
  
\bibitem{francis_g-2_lat_2014}
  A.~Francis {\it et al.} 
  arXiv:1410.7491.

\bibitem{Adler:1974gd}
  S.~L.~Adler,
  Phys.\ Rev.\ D {\bf 10} (1974) 3714.

\bibitem{Sachrajda:2004mi}
  C.~T.~Sachrajda and G.~Villadoro,
  Phys.\ Lett.\ B {\bf 609} (2005) 73, hep-lat/0411033.

\bibitem{Bedaque:2004ax}
  P.~F.~Bedaque and J.~-W.~Chen,
  Phys.\ Lett.\ B {\bf 616} (2005) 208, hep-lat/0412023.

\bibitem{deDivitiis:2004kq}
  G.~M.~de Divitiis, R.~Petronzio and N.~Tantalo,
  Phys.\ Lett.\ B {\bf 595} (2004) 408, hep-lat/0405002.

\bibitem{CLSwiki}  
  https://twiki.cern.ch/twiki/bin/view/CLS/WebIntro (2010).
  
\bibitem{CLSalgo}  
 http://luscher.web.cern.ch/luscher/DD-HMC/index.html
 
\bibitem{CLSpos}
 M. Marinkovic and S. Schaefer, PoS LATTICE2010 (2010) 031.  

\bibitem{latticespacing}
  S.~Capitani {\it et al.} 
  PoS LATTICE {\bf 2011} (2011) 145, arXiv:1110.6365.
  
\bibitem{Aubin:2012me}
  C.~Aubin, T.~Blum, M.~Golterman and S.~Peris,
  Phys.\ Rev.\ D {\bf 86} (2012) 054509, arXiv:1205.3695.

\bibitem{Lautrup:1971jf}
  B.~e.~Lautrup, A.~Peterman and E.~de Rafael,
  Phys.\ Rept.\  {\bf 3} (1972) 193.

\bibitem{Knecht:2003kc}
  M.~Knecht,
  Lect.\ Notes Phys.\  {\bf 629} (2004) 37, hep-ph/0307239.

\end{thebibliography}
\end{document}